# The assumptions that restrain us from understanding consciousness

Jaan Aru

jaan.aru@gmail.com

Institute of Computer Science, University of Tartu, Estonia

## Abstract

The science of consciousness has been successful over the last decades. Yet, it seems that some of the key questions remain unanswered. Perhaps, as a science of consciousness, we cannot move forward using the same theoretical commitments that brought us here. It might be necessary to revise some assumptions we have made along the way. In this piece, I offer no answers, but I will question some of these fundamental assumptions. We will try to take a fresh look at the classical question about the neural and explanatory correlates of consciousness. A key assumption is that neural correlates are to be found at the level of spiking responses. However, perhaps we should not simply take it for granted that this assumption holds true. Another common assumption is that we are close to understanding the computations underlying consciousness. I will try to show that computations related to consciousness might be far more complex than our current theories envision. There is little reason to think that consciousness is an abstract computation, as traditionally believed. Furthermore, I will try to demonstrate that consciousness research could benefit from investigating internal changes of consciousness, such as aha-moments. Finally, I will ask which theories the science of consciousness really needs.

Table of Contents

# Introduction: Have we made progress?

Have we made progress in consciousness science over the last decades? On the one hand, the answer to this question is a resounding 'Yes: we have!'. There is an established research tradition of consciousness. Editors are no longer reluctant to consider papers studying consciousness. Large grants are handed out to consciousness researchers. However, despite significant advances in consciousness science over the past few decades, fundamental questions about the nature of consciousness remain as elusive as ever.

We have seen the development of theories of consciousness, each offering distinct frameworks for understanding how consciousness might arise from neural processes (e.g., Dehaene et al., 2001; Dehaene & Changeux, 2011; Lamme, 2006; Lau & Rosenthal, 2011; Tononi, 2004; Tononi et al., 2016; overview of theories in Seth & Bayne, 2022). These theories have been the foundation of extensive research programs, with key papers on these topics accumulating thousands of citations, which is a testament to their influence and the attention they have received within the scientific community.

Yet, when scrutinized closely, the evidence supporting these theories often appears less concrete than one might expect. While there is a wealth of experimental data, much of it is, in fact, circumstantial, providing correlations between macro-scale markers of neural activity and conscious experience but falling short of actually explaining consciousness.

I would argue that consciousness science is at a crossroads, where existing theories may need to be reconsidered to achieve a deeper understanding of consciousness. These theories offer models, metaphors, and frameworks, and they give us a language to talk about consciousness, yet they do not penetrate the core mystery: how and why brain activity is accompanied by conscious experience. This realization should be humbling, prompting us to reconsider whether our current paradigms are sufficient or whether we need to search for new ways of thinking about consciousness. Perhaps we have not been able to take the next step because these theories offer a perspective on consciousness from a higher-level approach, i.e., they usually do not start from the biological details but rather from some (more or less) abstract ideas (e.g., Dehaene et al., 2001; Lau & Rosenthal, 2011; Tononi, 2004). The challenge, therefore, is not just to refine our existing theories but perhaps to rethink our assumptions, remaining open to new ideas that could one day offer a different path to explain consciousness.

Moreover, there is a sense in which we may find ourselves in a more challenging position today than we were two decades ago. Twenty years ago, the field was in a state of greater intellectual fluidity. There was an openness to radical ideas (for instance, the Integrated Information Theory, IIT, was just emerging, Tononi, 2004), a sense of exploration without the heavy weight of established theories. Today, however, the landscape has changed. The dominant theories of consciousness (including the IIT itself) have become so influential that they now exert a constraining effect on the field. For instance, over the last years, substantial grant funding was obtained to test competing theories of consciousness (Melloni et al., 2021), incentivizing many talented researchers to use their effort to contrast and compare these theories. On the one hand, the results have been another proof of the general success of consciousness science (Cogitate Consortium et al., 2025). However, at the same time, I would argue that this large-scale experiment demonstrated the constraining effects of the theories of consciousness, as these researchers spent their time and effort on these particular theories instead of developing new ideas. Hence, our current theories shape the questions we ask, the experiments we design, and the interpretations we consider plausible. This risks narrowing our collective imagination. By adhering too closely to established theories, we may limit our capacity to think outside the box, to explore ideas that do not fit within the prevailing paradigms. In this way, the very progress we have made over the past twenty years may paradoxically be holding us back, locking us into a way of thinking that could prevent us from making the next leap forward. Perhaps we should explore more and rely less on exploiting the existing theories.

Thus, the challenge now is to find a balance between building on the progress we have made and remaining open to the possibility that the true explanation of consciousness might require us to transcend our current theories altogether. While I am not arguing that this is the only solution, here I advance the idea that perhaps we could use the momentum to explore what other possibilities are out there.

The following section will start by taking a fresh look at the old problem of the neural correlates of consciousness. As we will see, the assumption has been that the pattern of spiking is a key *neuronal* correlate of consciousness. Here, I do not claim that everyone explicitly endorses this assumption. Rather, the point is that we have explicitly discussed this question too little, and hence, researchers do not even know that they could question this (implicit) assumption. We will hopefully see that the reasoning behind this assumption is rather superficial. We then address whether consciousness is a computation and conclude that perhaps consciousness cannot be reduced to straightforward, abstract computational processes. Further, I suggest that instead of studying *sensory* consciousness, much could be gained by studying internal changes of consciousness, such as aha-moments. Finally, we will ask which theories of consciousness the field still needs.

## Will the real NCCs please stand up?

One example of a clear and persistent open question in consciousness science revolves around the neural correlates of consciousness (NCCs). Although there is a wealth of literature exploring the NCCs, and numerous studies have identified potential links between neural activity and conscious experience, there remains little consensus on what constitutes the key correlates of consciousness (Aru et al., 2012; De Graaf et al., 2012). Researchers have identified various brain regions, patterns of neural firing, and specific types of neural activity that seem to correlate with conscious states, yet the picture remains fragmented and incomplete. This lack of agreement underscores the complexity of the problem and the challenges that still lie ahead.

One central problem is that we are used to thinking of phenomena that neatly map to some specific level of activity, for example, the level of neural spiking. However, it is possible that consciousness does not reside neatly within any single level of neural activity but rather emerges from the interactions across multiple scales (Storm et al., 2024). Even more, it is possible that the notion of a "level" is a neat abstraction but not something that Mother

Nature really cares about. In other words, perhaps there are no clearly distinguishable levels of activity in the brain.

## Is consciousness simply a very complex pattern of spiking?

Perhaps some readers would like to stop me right here and point out that surely we already know the relevant scale: it has to be the level of spiking activity of neural populations (Crick & Koch, 1990; Koch, 2004; Dehaene & Changeux, 2011; Tononi et al., 2016).

Spikes are the most conspicuous signals produced by neurons, representing rapid, all-or-nothing electrical impulses that travel along axons to transmit information. Because of their clear, measurable nature, spikes have become central to our understanding of how the brain processes information. In simple sensory systems, for example, research dating back to the first half of the 1900s demonstrated that the number of spikes a neuron generates corresponds to the intensity of a stimulus (Adrian, 1928). This suggests a straightforward "rate code" where the frequency of spikes encodes the strength of sensory input.

These types of spiking codes have been well-documented in sensory processing and other areas of neural computation. But does this mean they are also the mechanisms that underlie consciousness? Since the 1990s, this has been the leading assumption, perhaps best described by Crick and Koch in their seminal paper: "At any moment consciousness corresponds to a particular type of activity in a transient set of neurons that are a subset of a much larger set of potential candidates. The problem at the neural level then becomes: 1. Where are these neurons in the brain? 2. Are they of any particular neuronal type? 3. What is special (if anything) about their connections? 4. What is special (if anything) about the way they are firing?" (Crick and Koch, 1990)

This paper defined the research problem but also quite forcefully centered the search for the NCCs around neural firing patterns. While this might have seemed the most obvious physiological correlate at the time, it is essential to keep in mind that this is simply an assumption that deserves closer scrutiny.

In particular, this assumption that the level of spiking is the right level to understand consciousness may be more a product of convenience and the history of scientific progress than a reflection of the true nature of consciousness. Spikes simply are prominent, measurable, and were early on found to have clear correlations with certain perceptual experiences (e.g. Logothetis & Shall, 1989; Leopold & Logothetis, 1996) but none of these properties mean that spikes indeed underlie consciousness.

The problem with focusing on spiking patterns as the main correlate of consciousness is that these spikes, even when occurring in large neural populations, are not inherently bound into a coherent, unified entity, but consciousness is (Tononi, 2004, 2008; Bayne, 2012; Bachmann et al., 2020). Each spike is a discrete event, occurring independently within individual neurons, which raises the question of how these separate events could combine to produce the seamless, unified experience that characterizes consciousness. (Note that the reader might here say that spikes do not need to be combined, as consciousness could be an emergent property arising from spikes. However, as discussed in more detail below, the question then arises: why spikes, and not some other property of neural activity, then?).

This issue is highlighted by a thought experiment proposed by Albert Gidon and colleagues (Gidon et al., 2022), which presents a provocative challenge to the idea that spiking patterns alone can account for consciousness. Without explaining the specifics of the thought experiment, the core insight is that when we assume that spikes are the NCC, we will get into trouble. Namely, given that spikes are all separate events happening in individual neurons, we could theoretically scatter the neurons that generate these spikes relevant to any conscious percept without disrupting the pattern of spikes (i.e., if neurons 1 to n fire in a specific manner within 1 second, we could record this spiking, scatter the neurons and play exactly the same activity patterns back to the neurons in scattered neurons, see Gidon et al., 2022). Hence, if consciousness corresponds to the pattern of spiking, consciousness should be able to rely on spikes scattered (e.g., around the room or around the globe). In this scenario, the spatial arrangement of the neurons is irrelevant to the pattern itself; the same pattern of spikes could occur regardless of whether the neurons are closely packed or widely dispersed around the globe. If consciousness were solely a matter of these spiking patterns, this scattering should not make a difference, i.e., consciousness should persist unchanged. It might be that, in the end, it turns out that consciousness can be scattered, but for now, at least intuitively, the idea of scattered neurons having consciousness seems problematic. Consciousness feels like a unified, localized experience, not something that could be sustained by a scattered, disjointed network of neurons (Gidon et al., 2022).

This thought experiment serves as a critical reminder that there might be a fundamental problem with considering spiking patterns as the primary correlate of consciousness. It suggests that the relationship between spiking patterns and conscious experience may not be as straightforward as generally assumed. It is important to note that not all theories consider consciousness a simple byproduct of neural spiking. For instance, for the IIT, consciousness is related to the causal structure related to these spikes, hence, IIT also does not fall prey to the thought experiment (see Gidon et al., 2022).

While some might dismiss this thought experiment as purely speculative, it raises important questions that should not be ignored. It prompts us to reconsider whether spiking patterns, by themselves, can fully explain consciousness or whether consciousness might emerge from more complex neural processes that go beyond spiking.

## Consciousness and emergence

It has also been suggested that consciousness should be best understood as an emergent property. Generally, "emergence" refers to phenomena in which complex, coordinated behavior arises from the local interactions of simpler components, without the need for centralized control or long-range coordination (Anderson, 1972; Holland, 1998; Mitchell, 2009). Nonetheless, emergence can also describe situations where processes at a lower scale result in relatively simpler outcomes at a higher scale, underscoring that emergent phenomena may manifest as either greater organizational complexity or apparent simplification, depending on the context (Kim, 1999; Deacon, 2012). The fact that the macro-scale is relatively simpler and stable does not mean that it cannot have causal effects on the micro-level (Hoel et al., 2013).

In an emergent system, individual elements operate according to their own local rules, responding to both immediate inputs and broader, long-range signals. While each neuron might seem to act independently, their collective behavior gives rise to coordinated patterns that we perceive as organized and purposeful (e.g., Kelso, 1995; Singer, 2001; Varela et al., 2001; Di Volo & Destexhe, 2021). In this sense, coordination and coherence emerge naturally from the interactions within the system, without any need for a central controller.

If consciousness is indeed an emergent phenomenon, it suggests that our approach to studying it might need to shift. Instead of assuming that patterns of neural spiking underlie consciousness, we might focus on understanding the local rules governing neuronal interactions.Consider the analogy of a flock of birds in flight. The movement of the flock appears highly coordinated as if the birds are following a shared, central plan. However, we know that no such central coordination exists. Instead, each bird follows simple, local rules by adjusting its speed and direction based on the movements of its nearest neighbors. The overall pattern of the flock's movement emerges from these local interactions, without the need for any bird to know what the entire flock is doing.

Similarly, just as the flock's coordination does not require long-range communication or central control, the coordination in the brain underlying consciousness might also arise from local mechanisms. This means that although spikes are the primary means for long-range coordination, the mechanism underlying consciousness does not necessarily have to involve long-range coordination, and hence it does not have to involve spikes. In the brain, neurons communicate locally through various means, and while spikes are a prominent feature of neural activity, they are not the only form of communication. For instance, calcium waves are slow-moving signals that spread through the network of neurons and thus represent another form of neural interaction. These waves can modulate neuronal activity over longer timescales and across larger regions, potentially playing a role in coordinating neural activity in ways that spikes cannot achieve. Unlike spikes, which are brief and discrete, calcium waves provide a more continuous and diffuse form of communication, possibly contributing to the sustained and integrated nature of conscious experience. The argument here is not that calcium waves underlie consciousness, but rather to point out that there are other mechanistic levels or scales besides spiking activity that might be additional explanation candidates for consciousness. As said before, the focus on spikes might have been a historical artifact. What if, in the history of neuroscience, the measurement of calcium signals had predated the measurements of spiking activity?

Another possibility is that consciousness could arise from other neuron-intrinsic mechanisms. Inside neurons, complex biochemical processes govern how signals are processed and how neurons respond to inputs. Intracellular signaling pathways can modulate neural activity and synaptic strength over longer periods. These processes are more akin to the subtle, ongoing adjustments seen in flocking, where each bird's behavior is constantly influenced by its environment and neighbors. The critical point here is that while spikes are a key aspect of neural communication, they might not be the primary mechanism by which consciousness arises. Just as the flock's coordinated movement emerges from local interactions among individual birds, consciousness might emerge from a complex interplay of local neural mechanisms that go beyond simple spiking patterns.

This perspective suggests that researchers should broaden their focus when studying the neural correlates of consciousness. Instead of mainly investigating and modelling spiking responses, they should also consider other forms of neural communication and coordination. Other closely related fields have already done so. For instance, whereas for a long time, it was assumed that working memory is coded by persistent neural firing activity (Fuster, 2001), the field has by now established that part of working memory encoding happens by silent mechanisms within the synapses (e.g., Mongillo et al., 2008; Stokes, 2015; Panichello et al., 2024). By exploring these and other potential mechanisms, we might gain a deeper understanding of how consciousness emerges from the complex and dynamic activity of the brain, much like the coordinated movement of a flock arises from the simple, local rules followed by individual birds. However, we should also keep in mind that this simple analogy might be misleading: if consciousness were as simple as the flocking of birds, we would perhaps have already understood it by now. That said, the flocking behavior became tractable only once a large number of birds became trackable.

## Neural processes underlying consciousness do not have to be simple

All of the above is to say that perhaps we are not as close to understanding consciousness as we might have thought. While today's leading theories of consciousness provide valuable frameworks for understanding certain aspects of conscious experience, they may not capture the full biological and computational complexity of consciousness (Aru et al., 2023). Perhaps consciousness cannot be reduced to straightforward, easily understandable mechanisms.

The current theories of consciousness often operate at a high level of abstraction, focusing on the computational principles underlying consciousness without necessarily addressing the specific biological mechanisms involved. For instance, concepts like "information integration", "global workspace" or "recurrent feedback" are useful for framing discussions about consciousness, but they might gloss over the detailed neural processes that make these computations possible (Aru et al., 2023). These ideas are helpful for explaining in abstract and intuitive terms how different regions of the brain might work together to produce conscious experience. However, they abstract away the complexity of biological processes that take place at the cellular and molecular levels. If these levels play any crucial roles in the emergence of consciousness, we have abstracted away the key components that a true theory of consciousness requires.

To truly understand consciousness, we may need to move beyond the current level of abstraction and develop theories that embrace the complexity of the biological processes involved. This could entail exploring how different levels of neural processing (from molecular signaling pathways to large-scale brain networks) interact to produce consciousness. For instance, one could study how specific molecular processes, such as neurotransmitter release or ion channel dynamics, influence neural computation at the cellular level, and how these computations scale up to influence network dynamics and ultimately conscious experience (Aru et al., 2020; Suzuki & Larkum, 2020; Storm et al., 2024). Again, we do not know what level of complexity is needed to understand consciousness, but we should be aware that it might be more complex than our present theories suggest.

## Is consciousness a computation?

But the problem runs deeper. At the heart of many popular theories of consciousness lies a foundational assumption: computational functionalism. (This is not true for all theories. For instance, see Findlay et al., 2025 for a refutation of computational functionalism from the perspective of IIT). According to computational functionalism, consciousness arises from specific computational processes, independent of the physical substrate in which these processes occur (Putnam, 1967; Colombo and Piccinini, 2023). In essence, this view asserts that the substrate (biological, silicon-based, or otherwise) is irrelevant as long as the requisite computational processes are performed. Consider, for example, the function of addition: whether performed by a human brain, an abacus, or a computer, the result remains fundamentally the same as long as the operations adhere to the underlying principles of arithmetic. Computational functionalism is tied to the concept of "multiple realizability," which suggests that mental and experiential states can emerge in any system capable of executing the appropriate computations. One of the consequences of computational functionalism is that if artificial intelligence systems execute the "right" kinds of computations, they could potentially achieve consciousness (Butlin et al., 2023).

However, the validity of computational functionalism regarding consciousness hinges on the existence of computational patterns that have two properties. First, they should be able to arise from different underlying physical systems while maintaining functional equivalence at a higher level. Second, for a researcher to implement consciousness in other systems, these computational patterns first have to be recoverable and understandable in the brain without referring to lower-level details.

This implies that computational functionalism relies on the idea that some computations can be shielded from what happens at the lower level, i.e., that a higher-order state can be understood without taking into account the lower-level processes (see Barnett & Seth, 2023; Rosas et al., 2024 for more formal descriptions). In short, computational functionalism claims that information processing can be coherently described at higher levels without needing to track low-level physical details.

However, as we saw above, the assumption that consciousness arises from emergent computational patterns such as spiking likely oversimplifies the brain's structural and functional complexity. Emergence, traditionally conceived, implies distinct levels—low-level interactions giving rise to high-level properties that maintain causal closure. However, it could be argued that the brain does not exhibit clear-cut levels of organization in this sense. Instead, neural processes might involve deeply interwoven dynamics where no sharp boundary separates low-level and high-level activities (Milinkovic et al., 2025; Milinkovic & Aru, forthcoming).

Without well-defined levels, the abstraction of computational equivalence might miss critical features intrinsic to the brain's operation, potentially undermining the principle of multiple realizability. As Rosa Cao (2002) has said, "... we have no theoretical guarantee that any such efficient, intermediate level of description exists". The only guarantee we have is that our brains generate consciousness.

The problem partly lies in the fact that brains do compute and can support computational processes (Colombo and Piccinini, 2023). However, from this fact, one should not conclude that consciousness is also a computation in a traditional sense. It can be shown with a computational simulation that one can generate circumstances where computation is degenerated, while neural processes work exactly the same as before (Gidon et al., 2025). In this case, does consciousness change? In other words, does consciousness vary with the degeneration of computation, or does it stay intact hand-in-hand with neural processes? A computational functionalist has to bite the bullet and think that consciousness varies while there are no changes to neural activity (Gidon et al., 2025). This stops being strange the moment we realize that consciousness is not an abstract computation.

## Matter matters for the mind

If consciousness is not an abstract computation, then what is it? One possibility is that consciousness does not merely emerge from the brain's computational capacities but rather from the deeply rooted constraints of evolution and biology. This perspective draws on the principle of generative entrenchment, originally outlined by Wimsatt (1986), and more recently discussed by Cao (2022) and Seth (2025). Generative entrenchment posits that lower-level foundational features in a complex system are deeply interwoven with high-level processing. In the brain, this means that complex biological processes become deeply embedded and specialized over evolutionary times and serve as scaffolds for the emergence of higher-level functions, including aspects of mind and subjective experience.

To see the stark contrast to biology, consider artificial neurons. They are just a piece of computer software, a piece of code (Aru et al., 2023). In biology, on the contrary, matter matters. The specific physical organization within the cell is what allows DNA to be transcribed to RNA; the molecules and their shape enable the translation of RNA to proteins. One can describe these processes computationally, but no proteins will emerge from that computation because the true translation process requires specific physical processes. What if the computations related to consciousness similarly require specific physical processes that happen within the brain?

Another way to say this is to say that *part of the computation is done by the physical matter itself*. The computation of RNA to proteins is partly describable abstractly in the code of how specific parts of RNA correspond to specific proteins but a key part of the computation is done by the physical processes within the cell. In short, abstract computation is only a (tiny) part of the game. Similarly, there might be abstract computations related to consciousness that can be implemented in digital computers. But the other part of these computations might be physical, currently done only in brains.

The physical part of computation is potentially lethal for multiple realizability because the more the abstract computations depend on the specific machinery, the fewer possibilities there are for realizing this computation elsewhere. The larger the physical part is, the more lethal it is to computational functionalism. When it comes to consciousness, we have no clue how big the physical part is. But there is no reason to think that it should be nonexistent.

Biology is highly specific. These physical properties significantly constrain what types of processes can be run on them. As Rosa Cao (2022) has pointed out very nicely: "Once a few simple molecules or components are fixed (say water or oxygen), they then constrain the components that can smoothly interface with them in the right ways, which pose their own constraints in turn on other components, until possible realizations of the whole system have been clamped down to options not significantly different from our own biological one."

This does not mean that consciousness can never be understood at the level of computations. Eventually, we might be able to crack the code and implement the physical part of computations also *in silico*. However, it should be clear that these are computational processes beyond our current capabilities or imagination. Until that day, it does not seem reasonable to assume that consciousness can be captured as an abstract computation. Matter matters for the mind.

## Beyond perception: Sudden shifts in consciousness

In 2011, our paper on "Distilling the neural correlates of consciousness", which predominantly focused on vision, had a relatively tough reviewer. In addition to explaining why they did not like the paper, the reviewer wrote: "In addition, of course there are conscious events that are not stimulus-triggered like our typical vision experiments. Some notable work on that has been done by Jung-Beemann and colleagues, in the context of intuitive problem-solving and the traditional "Aha!" experience."

Back then, I didn't think much of it, and we argued against including this in our manuscript. Yet, now, perhaps because this critique lingered somewhere in my brain, I'm slowly coming back to this idea that besides these "stimulus-triggered" experiments, there are indeed interesting conscious experiences that happen internally. The key problem with stimulus-triggered experiments is that it is practically quite impossible to disentangle the true NCCs from the processes that precede and follow it (Aru et al., 2012; De Graaf et al., 2012). Thus, we need to study other types of experiences. And one of them is, in fact, the "aha!" experience (Jung-Beeman et al., 2004; Danek & Wiley, 2024; Tulver et al., 2023, 2025).

"Aha!" experience can, of course, sometimes be stimulus-triggered, for instance, if you see a Mooney image or an anagram and immediately solve it. But more striking and interesting for consciousness researchers are the cases where the initial presentation of the stimulus does not elicit the solution. During these cases, while behaviorally nothing happens, an internal reorganization, restructuring (Danek & Wiley, 2024; Tulver et al., 2023) can take place and the solution might suddenly appear in consciousness.

Here, then, a solution comes from the inside and pops into consciousness, due to some neural processes we still do not understand. There are, of course, plenty of EEG and fMRI works on the neural correlates of "aha!" experiences (Jung-Beeman et al., 2004; Kounios & Beeman, 2014); however, as discussed above, the EEG and fMRI level studies are simply not an adequate level to explain consciousness. Thus, to understand the neural mechanisms of "aha!" experiences as conscious experiences, we need to dig deeper.

Next to the lab-induced insights, there are also the more profound "aha!" moments often described as mystical or transcendent experiences. Historically, William James (1902) offered detailed accounts of such states in *The Varieties of Religious Experience*. In these transformative moments, an insight or realization seems to flood the entire psyche, leading to a sense of radical reorganization of one's worldview (see Miller & C'de Baca, 2001, for more modern examples).

In insight, this manifests as a feeling that disparate pieces of information have suddenly come together in a coherent whole (Ohlsson, 1984; Tulver, 2023). In mystical experiences, it is expressed as a dissolution of the boundaries between self and world, and a feeling of connection to something larger than oneself. Consciousness changes. For example, consider this experience by Ruben Laukkonen (2024): "As minutes passed I felt as if my awareness was increasing. I felt more sensitive and open, as if I was taking in more of the moment than I normally did, as if I had taken off some dirty old sunglasses and my peripheries were coming alive. It was as if a grey cloud that permeated everything had finally cleared from the air. It was as if I was 'gaining' consciousness, whatever that substance is, which we all intuitively feel we have. The capacity to experience—if that is what consciousness is—seemed to be expanding and able to hold together more of what is, becoming richer and more refined."

These mystical experiences and major "aha!" moments illustrate a fundamental property of consciousness that has not been at the center of consciousness research: our consciousness can be fundamentally reorganized quite abruptly. This implies that any neural mechanism underlying consciousness also has to have the capacity to rearrange correspondingly. In other words, the fact that such experiences exist puts constraints on the possible mechanisms of consciousness.

The crucial aspect of any explanation of "aha!" experiences is the abrupt shift from not having an insight to having one. So the key challenge is to figure out the neural mechanisms of how such a rapid, discrete transition can occur internally, without any additional external input. What could such restructuring look like in terms of cellular and circuit dynamics? And how could these local neural events yield a phenomenal experience of insight?

We do not have a great explanation for such changes in consciousness. In consciousness science, the phenomenon most closely related to insight may be the concept of *ignition*, which involves a nonlinear change in neuronal activation. A recent study by Klatzmann and colleagues (2024) employed a mesoscale, connectome-based dynamical model of the macaque cortex, incorporating NMDA, AMPA, and GABA receptors. Their results indicate that *ignition* critically depends on the activation of NMDA receptors in local excitatory circuits.

This NMDA-dependent ignition can be tied directly to how insights occur. When a person is attempting to solve a puzzle, multiple neuronal groups may encode partial aspects of the problem, without forming a coherent representation that corresponds to a solution. During this subthreshold or "pre-solution" state, many local circuits are effectively "trying out" different configurations, but none are stable enough or sufficiently complete to yield a conscious "aha!" moment. Meanwhile, factors such as short-term synaptic plasticity, neuromodulation, and local cellular properties gradually shape the likelihood that certain nodes in the network will become more dominant.

Crucially, NMDA receptors can play a pivotal role in determining when and how these local circuits transition from transient, subthreshold activity to a robust, self-sustaining pattern. Through the NMDA receptor's voltage-dependent and longer-lasting conductance, even small changes in synaptic strength or neuronal excitability can tip the entire system toward a new attractor state—an emergent spiking pattern that brings together the right combination of neuronal ensembles for the solution (Tulver et al., 2025). This abrupt onset of a stable, coherent firing pattern might reflect the qualitative leap that feels immediate and revelatory to the person experiencing it.

As should be clear from the discussion above about the NCCs, we do not know whether this story is adequate or not. It might be that the suddenness of insight can be understood as the neural system shifting from an incomplete representation to a firmly stabilized, self-reinforcing state. But this is just speculation and there might be other, yet unknown mechanisms responsible for the insight phenomenon and conscious experience in general. Regardless, the main message is that abrupt activations, such as insight moments, provide a window into the neural underpinnings of consciousness as they illustrate that consciousness often involves discrete state transitions even without the stimulus. The fact that individuals experience a sudden shift in phenomenological clarity indicates that the neural mechanisms of consciousness must be able to spontaneously reorganize. By investigating these abrupt state transitions at cellular and circuit levels, we may uncover broader principles of how consciousness emerges from the interplay of distributed, multi-level, self-organizing processes across the brain.

## What type of theory of consciousness do we still need?

There is a plethora of theories of consciousness (Seth & Bayne, 2022; Kuhn, 2024), hence it seems preposterous to propose that we need yet another theory. However, it is possible that the current theories of consciousness do not cover the whole landscape of potential theories.

In particular, as mentioned above, existing theories often remain relatively abstract and do not fully address the biological details that might give rise to conscious experience (Aru et al., 2020; Storm et al., 2024). This limitation is especially salient given that consciousness is a biological phenomenon, and biology is subject to principles such as "entrenchment "(Cao, 2022; Seth, 2024).

In the second part of this chapter, it was discussed how insight and spontaneous restructuring demonstrate that consciousness is not merely reactive to external stimuli—it emerges endogenously from within the brain. Although the idea that consciousness can arise in the absence of external stimulation is straightforward, many foundational theories in cognitive science were originally constructed around stimulus-driven processes (Dehaene & Naccache, 2001; Lamme, 2004; Mashour et al., 2020). In these accounts, once a stimulus is sufficiently strong or presented for a long enough duration, it can enter consciousness. While such theories have been used to explain internal processes (Moutard et al., 2015) and can be extended to explain internally generated phenomena (such as imagination, dreaming, hallucinations, or the "aha!" moments of sudden insight) they often do so by positing extra mechanisms or assumptions that do not stem naturally from their core frameworks.

By contrast, some accounts, such as Integrated Information Theory (IIT), propose that consciousness arises from intrinsic properties of certain physical systems (Tononi, 2004; Oizumi, Albantakis, & Tononi, 2014). These intrinsic properties purportedly endow conscious systems with inherent meaning and unified experience. Hence, IIT is in a position to explain internally generated phenomena (Mayner et al., 2024).

However, some researchers find that IIT, in its current formulations, makes too specific theoretical commitments that can be difficult to test empirically or that appear too rigid for some lines of research (Bayne, 2018; Merker et al., 2022). Nonetheless, IIT usefully highlights that consciousness might be best explained by theories focusing on *intrinsic* rather than strictly *extrinsic* or *functional* aspects of cognition. The key takeaway here is not that IIT is the only approach but that *we need more theories like IIT that are not IIT*. Some of IIT's conceptual tools, particularly its emphasis on intrinsic meaning and the structural underpinnings of experience, do not constitute an "intellectual property" of IIT alone. Rather, they can be productively adapted, refined, and used in alternative theoretical frameworks.

One especially pressing gap is the lack of a formal, computational theory of *biological naturalism*, i.e., the view that consciousness is a natural biological phenomenon irreducible to purely functional or computational descriptions (Searle, 1992; Seth, 2025). Although the phrase "computational theory of biological naturalism" may sound contradictory, such a theory would aim to clarify which aspects of biological organization and entrenchment can be captured in computational or algorithmic terms, and which aspects currently remain elusive. As explained above, our current theories might lack crucial details about neurobiology. Hence, we can first improve our computational view of consciousness by simply adding these computational details to our models and theories (see Munn et al., 2023; Klatzmann et al., 2024; Whyte et al., 2025 for recent examples). Second, we can better state, in computational terms, which processes are still left out of our models of consciousness, for instance, the metabolic constraints of biological matter (e.g., Milinkovic & Aru, forthcoming). Only by making explicit what current computational tools can and cannot model will we be able to delineate the frontiers of our understanding and make meaningful progress in explaining consciousness in all its biological complexity (and beauty).

## The discoveries we cannot predict

We are constrained by our theories of consciousness. Our experiments, our models, and even the language we employ are all shaped by existing theoretical frameworks. This is not to say that established theories are without merit; indeed, they have driven significant progress in our understanding of the neural correlates of consciousness and have guided empirical research over the past several decades. And yet, our thinking is handicapped by thinking about consciousness in the terms and theoretical constructs of these theories.

The moment we allow for the possibility that consciousness operates in ways not yet captured by our existing theoretical or computational models, we open the door to discoveries we cannot yet anticipate. In the end, acknowledging the limits of our current theories is not an admission of defeat but an invitation to explore the unknown. Real progress in consciousness research will hinge on our willingness to question, refine, and abandon assumptions.

In the end, I wish we were more humble. Consciousness is probably more complex than our simple theories. That does not mean that we have to understand all the details of its neural machinery. But we have to remain open to the possibility that the true nature of consciousness could emerge from the very complexity that makes it so difficult to pin down.

**Acknowledgments**

I am grateful to Talis Bachmann, Albert Gidon, Alex Maier, Lucia Melloni, Borjan Milinkovic, Umberto Olcese, Kadi Tulver, Raul Vicente, Christopher Whyte and Zefan Zheng for their helpful comments. This work was supported by the Estonian Research Council grant PSG728 and the Estonian Centre of Excellence in Artificial Intelligence (EXAI), funded by the Estonian Ministry of Education and Research.